\documentstyle[twocolumn,epsf]{jpsj}

\catcode`\@=11

\def\simle{\mathrel{\mathpalette\@versim<}}   
\def\simge{\mathrel{\mathpalette\@versim>}}   
\def\@versim#1#2{\lower2.5pt\vbox{\baselineskip0pt \lineskip-.5pt
   \ialign{$\m@th#1\hfil##\hfil$\crcr#2\crcr\sim\crcr}}}

\catcode`\@=12

\def\runtitle{
}
\def\runauthor{Yukitoshi {\sc Motome} and Nobuo {\sc Furukawa}}

\def\sf{\rm}

\title{
Anomaly in Spin Excitation Spectrum of \\
Double-Exchange Systems with Randomness
}

\author{
Yukitoshi {\sc Motome} and Nobuo {\sc Furukawa}$^{1}$ 
}
\inst{
Institute of Materials Science, University of Tsukuba,
Tsukuba, Ibaraki 305-0006 \\
$^{1}$Department of Physics, Aoyama Gakuin University, 
Setagaya, Tokyo 157-8572
}
\recdate{
}

\abst{
Spin excitation spectrum of the double-exchange model 
is studied in the presence of randomness.
Spin wave approximation in the ground state shows that
the randomness significantly modifies 
the spectrum from the  cosine-like one in the pure system
to that with anomalies
such as broadening, anti-crossing and gap opening.
The origin of anomalies is speculated to be modulation of 
effective ferromagnetic coupling by the Friedel oscillation.
These anomalies qualitatively reproduce the spin excitation spectrum
in colossal magnetoresistance manganites
whose Curie temperatures are relatively low.
Our results suggest that randomness control is an important notion
to understand effects of the A-site substitution
which has previously been understood as the bandwidth control.
}

\kword{
colossal magnetoresistance manganites,
double-exchange model, randomness, spin dynamics,
A-site substitution, bandwidth control
}

\begin{document}
\sloppy
\maketitle


One of the most important issues of
colossal magnetoresistance (CMR) manganites AMnO$_{3}$
is to understand effects of the A-site substitution.
\cite{Ramirez1997}
Generally, the A-site substitution has two different aspects.
One is the carrier doping
which is the substitution by A ions with different valences.
The other is the ionic radius modification
which is the A-site substitution
with ions having the same valence but different ionic radius.
In the latter, the averaged radius of A ions
$\langle r_{\rm A} \rangle$ affects
length and angle of Mn-O-Mn bonds.
This leads to a change of effective transfer energies
between Mn sites.
Therefore, it is widely accepted that
the ionic radius modification gives a bandwidth control.
Owing to a wide potential of these A-site substitutions,
manganese oxides exhibit rich physics including the CMR phenomena.

Among various compounds AMnO$_{3}$,
recent developments in both experiments and theories have been revealed that
La$_{1-x}$Sr$_{x}$MnO$_{3}$ (LSMO) at $x \simeq 1/3$
is a canonical double-exchange (DE) systems.
\cite{Furukawa1999a,Motome2000}
The DE model
\cite{Zener1951}
can explain
magnetic and transport properties in this compound quantitatively.
For instance, the experimental values of  Curie temperature $T_{\rm C}$
agree well with the theoretical estimate.
\cite{Motome2000}
Thus, the LSMO compound provides a well-established starting point
to examine effects of the A-site substitutions.

Starting from LSMO as a reference,
we focus on the ionic radius control by the A-site substitutions.
This causes a systematic change in transport and magnetic properties.
For instance, the value of $T_{\rm C}$ decreases systematically
for compounds with smaller $\langle r_{\rm A}\rangle$
and hence with `narrower-bandwidth'  such as
Pr$_{1-x}$Sr$_{x}$MnO$_{3}$ and La$_{1-x}$Ca$_{x}$MnO$_{3}$ (LCMO).
Hereafter, we call these A-site substituted compounds
`low-$T_{\rm C}$ manganites'
whereas we call the compounds such as LSMO,
which show relatively high $T_{\rm C}$, `high-$T_{\rm C}$ manganites'.
An empirical relation between $T_{\rm C}$ and $\langle r_{\rm A}\rangle$
 has been proposed experimentally.
\cite{Hwang1995}
The systematic change has been considered
to be primarily due to the bandwidth change by the A-site substitution.

However, the picture of the bandwidth control
has been doubted by some experimental results.
One is a quantitative comparison between $T_{\rm C}$ and the bandwidth
estimated from $\langle r_{\rm A}\rangle$.
The change of $T_{\rm C}$ is much larger than that of the bandwidth.
For instance, from LSMO to LCMO, $T_{\rm C}$ decreases about 30\%
while the estimated bandwidth decreases less than 2\%.
\cite{Radaelli1997}
This large decrease of $T_{\rm C}$ cannot be explained
by the bandwidth control in the DE theory
in which $T_{\rm C}$ should scale to the bandwidth.
An extra degree of freedom as a hidden parameter 
seems to be  necessary to explain such a change.

In this Letter,
we  examine a role of the randomness in the A-site substitution.
In general, oxides are known to be far from perfect crystals.
Moreover, in manganites, 
randomness is inevitably introduced by the A-site substitution
since the compounds are solid solutions of different A ions.
Random distribution of A ions induces
a charge and structural disorder in the system, which introduces
randomness in the electronic Hamiltonian.

Importance of the randomness effects has been discussed 
both experimentally and theoretically.
One is that $T_{\rm C}$
is much affected not only by the averaged radius of A ions
but also by the standard deviation of the ionic radii.
A linear decrease of $T_{\rm C}$ as a function of the standard deviation
has been reported.
\cite{Rodriguez-Martinez1996}
Another experimental fact is a correlation
between the residual resistivity and $T_{\rm C}$.
The A-site substitution from high-$T_{\rm C}$ to low-$T_{\rm C}$ manganites
causes a systematic increase of the residual resistivity
even in single crystals.
\cite{Coey1995,Saitoh1999}
These results clearly indicate that 
it is necessary to take account of effects of the randomness
in the A-site substitution.
Recently, effects of the randomness have been discussed in the DE model.
\cite{Mazzaferro1985,Allub1996,Li1997,Sheng1997a}
A significant decrease of $T_{\rm C}$ by the randomness has been predicted.
\cite{Mazzaferro1985,Allub1996,Narimanov2000,Letfulov2001,Auslender2001}

Let us now consider another important result of 
the ionic radius modification which
is the systematic change in the spin excitation spectrum.
High-$T_{\rm C}$ manganites show a cosine-like spin-wave dispersion
\cite{Perring1996}
which is well described by the DE theory quantitatively.
\cite{Furukawa1996}
On the other hand, in low-$T_{\rm C}$ manganites,
the spin excitation deviates from this behavior considerably,
and shows some anomalies such as
softening, broadening, anti-crossing and gap-opening.
\cite{Hwang1998,Vasiliu-Doloc1998,Dai2000,Biotteau2001}
The DE theory also fails to explain these anomalies in the spin dynamics.

In this Letter,
 we study effects of the randomness on the spin dynamics in the DE model.
The spin excitation is calculated in the ground state by the spin wave approximation.
We find clear deviation from the cosine-like spin-wave dispersion
in the pure DE model and some anomalies in the spin excitation spectrum.
Through comparisons with the experimental results 
in low-$T_{\rm C}$ manganites,
our results strongly suggest that 
the randomness gives an important effect of the A-site substitution.


As a model for the A-site substituted manganites,
we consider the DE model with randomness.
The Hamiltonian is written by
\begin{equation}
{\cal H} = - t \sum_{\langle ij \rangle, \sigma}
( c_{i \sigma}^{\dagger} c_{j \sigma} + {\rm h.c.} )
- \frac{J_{\rm H}}{S} \sum_{i} \vec{\sigma}_{i} \cdot \vec{S}_{i}
+ \sum_{i \sigma} \varepsilon_{i} c_{i \sigma}^{\dagger} c_{i \sigma},
\label{eq:H}
\end{equation}
where 
the first two terms denote the DE Hamiltonian
\cite{Zener1951}
which consists of the nearest-neighbor hopping
and the Hund's-rule coupling between itinerant electrons 
and localized spins with a magnitude $S$.
These two terms quantitatively describe 
physical properties in high-$T_{\rm C}$ manganites
including the spin excitation.
\cite{Furukawa1999a,Motome2000,Furukawa1996}
The last term denotes the on-site randomness.
In this work, for simplicity, we incorporate
the randomness by the A-site substitution
as the on-site random potential
although it may cause various other effects,
for instance, a randomness in transfer integrals.
Here we consider the binary-alloy type randomness,
that is, $\varepsilon_{i}$ takes $\pm W_{\rm imp}/2$
in equal probability in each site.
We confirm that the results are independent of
the type of randomness qualitatively.
Although 
it is difficult to determine the actual
 magnitude of the randomness in manganites,
the value of $W_{\rm imp}$ is roughly estimated to be 
the same order of magnitude as the bandwidth in low-$T_{\rm C}$ compounds.
\cite{Coey1995,Pickett1997}
As an energy unit, we use the half bandwidth
at the ground state  for $W_{\rm imp} = J_{\rm H} = 0$,
i.e., $W = 6t = 1$ in three dimensions.

We study the spin excitation of the model (\ref{eq:H}) 
using the spin wave approximation.
We consider a  perfectly spin-polarized ferromagnetic 
ground state $\vec{S}_{i} = (0,0,S)$,
and calculate the one-magnon excitation spectrum 
 within the
lowest order of $1/S$ expansion.
In the absence of impurities, 
the formulation  is given  in Ref.~\citen{Furukawa1996}.

Here we extend the method to the case with randomness.
For a given configuration of the quenched randomness $\{\varepsilon\}$
on a finite size cluster,
we explicitly  diagonalize the Hamiltonian matrix
at the ground state. We have
\begin{equation}
  {\cal H} \tilde\varphi_{n\sigma}(i) 
= (\tilde E_n - \sigma J_{\rm H}) \tilde \varphi_{n\sigma}(i),
\end{equation}
where $\tilde \varphi_{n\sigma}$
 is the $n$-th wavefunction for the spin $\sigma$
electron.
Since  wavefunctions are given  in real space,
it is straightforward to formulate  Green's functions
of electrons and spin wave 
using the real-space representation.

\begin{figure}
\epsfxsize=5.5cm
\centerline{\epsfbox{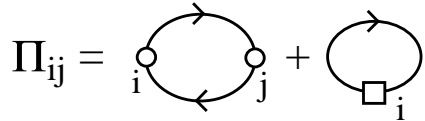}}
\caption{
Spin wave self-energy in the lowest order of the $1/S$ expansion.
See Ref.~\citen{Furukawa1996} for details.
}
\label{fig:SE}
\end{figure}

Electron Green's function is given by
\begin{equation}
\tilde G_{ij,\sigma}(\omega)
= \sum_n \frac{ \tilde \varphi_{n\sigma}(i) \tilde  \varphi_{n\sigma}^*(j)} 
{ \omega - (\tilde E_n - \sigma J_{\rm H} - \mu) + {\rm i}\eta\,{\rm sgn}
 \omega}.
\end{equation}
Following Ref.~\citen{Furukawa1996},
spin wave self-energy is
calculated from electron's spin polarization function
depicted in Fig.~\ref{fig:SE}.
In the limit of $J_{\rm H}/t \to \infty$, which is realistic in manganites,
we obtain 
\begin{eqnarray}
\tilde \Pi_{ij}(\omega)
&=&\frac{1}{2SN} \sum_{mn} f_{n\uparrow}
 \tilde \varphi_{n\uparrow}(i) \tilde  \varphi_{n\uparrow}(j)^* 
 \tilde \varphi_{m\downarrow}(j) \tilde  \varphi_{m\downarrow}(i)^*
\nonumber \\
&&\qquad\qquad \times \ 
(\tilde E_{m} - \tilde E_{n} - \omega),
\end{eqnarray}
where $N$ is the number of lattice sites and
$f_{n\uparrow}$ is the fermi distribution function for up-spin states.
Spin wave Green's function is given by 
 $\tilde D_{ij}(\omega) 
= (\omega - \tilde \Pi_{ij}(\omega) +{\rm i}\eta)^{-1}$.
Spin wave excitations are obtained from the poles
of the Green's function.
Since $\tilde \Pi_{ij} \propto 1/S$,
positions of poles are at $\omega \sim O(1/S)$.
Then, as discussed in Ref.~\citen{Furukawa1996},
spin wave excitations within the lowest order of $1/S$
expansion can be determined from the
static part of the self-energy $\tilde \Pi_{ij}(\omega=0)$.

\vspace{-4mm}

The spectral function for spin wave is defined by
\begin{equation}
 \tilde A(\mib{q},\omega) = 
-\frac1N \sum_{ij} \frac1{\pi}{\sl Im}\tilde D_{ij}(\omega)
\exp\left[{\rm i}\mib{q} \cdot (\mib{r}_{i} - \mib{r}_{j})\right] .
\end{equation}
Using the eigenvalues $\tilde \omega_l$
and eigenvectors $\tilde \psi_{l}(j)$ of $\tilde \Pi_{ij}$, 
\begin{equation}
\sum_j \tilde \Pi_{ij}(\omega={0})\cdot \tilde \psi_{l}(j)
 = \tilde \omega_{l} \tilde \psi_{l}(i),
\end{equation}
the matrix inversion of $(\omega - \tilde \Pi_{ij})$ 
is obtained.
We have
\begin{eqnarray}
\label{eq:A_epsilon}
\tilde A(\mib{q},\omega)
&=& \sum_{l} \tilde A_l({\mib{q}}) \delta(\omega - \tilde \omega_{l}),
\\
\tilde A_l({\mib{q}}) &=& \frac{1}{N} \Bigl| \sum_{j}
\tilde \psi_{l}(j) \exp({\rm i}\mib{q}\mib{r}_{j}) \Bigr|^2.
\end{eqnarray}

Finally, we average $\tilde A(\mib{q},\omega)$
for random configurations of the potential $\{\varepsilon\}$
to obtain the spectral function $A(\mib{q},\omega)$.

\begin{figure}
\epsfxsize=7cm
\centerline{\epsfbox{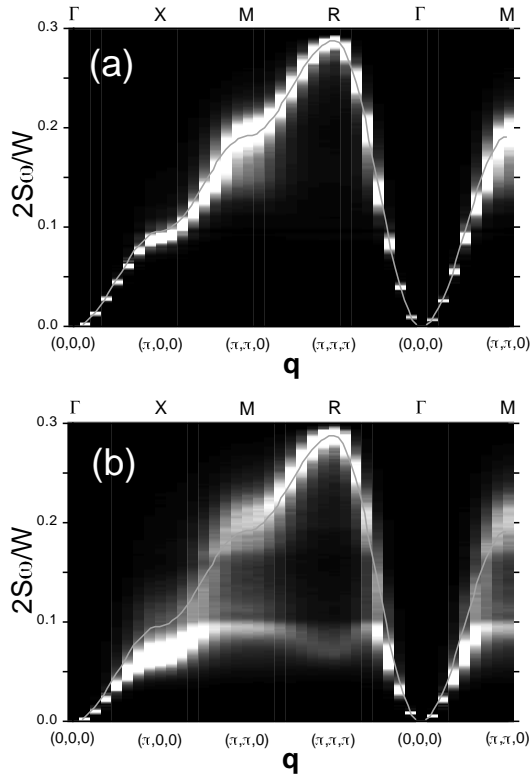}}
\caption{
Spin excitation spectra at $x=0.3$ in three dimensions 
for (a) $W_{\rm imp}/W = 0.2$ and (b) $W_{\rm imp}/W = 0.5$.
The intensity is shown by gray scale along the vertical axis.
The discreteness along the horizontal axis comes from
the fact that the calculations are performed on finite-size clusters.
The gray curves are the spin wave dispersion in the pure case.
}
\label{fig:3D}
\end{figure}

Figure~\ref{fig:3D} shows the spectral function for spin excitations
at $x=0.3$ in three dimensions.
Here, $x$ is the hole density in model (\ref{eq:H}).
System size $N$ is $16 \times 16 \times 16$ with the periodic boundary conditions
and the random average is taken for $24$ different configurations
of the random potential for each value of $W_{\rm imp}$.
For small values of $W_{\rm imp}$,
the spectrum is similar to the cosine-like dispersion for $W_{\rm imp} = 0$
(gray curve)
although it shows some broadening.
However, when the value of $W_{\rm imp}$ becomes moderate,
some anomalies appear conspicuously in the spectrum.
Apart from the large broadening,
a remarkable feature is a branching which is clearly seen, for instance,
along the ${\rm \Gamma}$-M and ${\rm \Gamma}$-R lines for $W_{\rm imp}/W = 0.5$.
Near the zone boundary,
extra branches with considerable weight appear
well below the cosine-like branch.
If one follows only the extra lower branch,
the spin excitation spectrum appears to show a significant softening.

\vspace{-4mm}

\begin{figure}
\epsfxsize=7.5cm
\centerline{\epsfbox{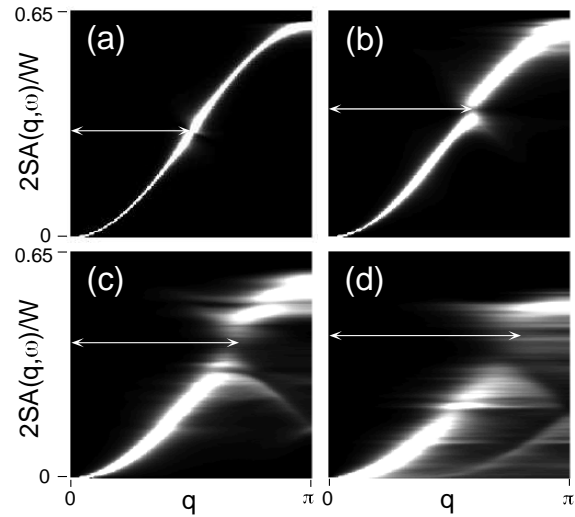}}
\caption{
Spin excitation spectra in one dimension for
(a) $x=0.5$, (b) $x=0.4$, (c) $x=0.3$ and (d) $x=0.2$.
The arrow in each figure indicates the magnitude of $k_{\rm F}$.
}
\label{fig:1D}
\end{figure}

In order to consider the origin of the anomalies,
we perform detailed calculations in one dimension.
We take $N = 256$ and the random average for $500$ configurations.
Figure~\ref{fig:1D} shows the results for different values of $x$
at $W_{\rm imp}/W = 0.5$ ($W = 2t$ in one dimension).
The precise calculations in one dimension reveal
the detailed structure of the anomalies;
an anti-crossing structure with a gap opening,
shadow bands and a significant broadening.
These remarkable features become conspicuous for the values of $x$
far from the quarter filling $x=0.5$.

An important point in Fig.~\ref{fig:1D} is
that the anomalies appear near the Fermi wave number $k_{\rm F}$
which is given by $\pi (1-x)$ in one dimension.
This indicates that
the anomalies come from a singularity of $2 k_{\rm F}$ momentum transfer
between $q \simeq k_{\rm F}$ and $q \simeq -k_{\rm F}$.
There are also small anomalies at
$q \simeq \pi - k_{\rm F}$ and $- \pi + k_{\rm F}$.
We note that these anomalies can be seen even in the case with 
a single impurity.
Namely, positions of anomalies in $q$-space
scale with $k_{\rm F}$, while the intensities of the anomalous parts
depend on impurity strength.
From these observations, it is clear that
the origin of the anomalies is the fermionic responses
of the itinerant electrons to impurities.

\vspace{-4mm}

We speculate that randomness induces modulation of
effective exchange coupling due to the Friedel oscillation.
The random potential causes the $2 k_{\rm F}$ singularity
in the charge channel, which
appears in the effective ferromagnetic coupling
through the magnon-electron interaction.
Scattering of magnons by the oscillating components of the exchange couplings
creates the anti-crossing structure.
The Friedel oscillation occurs universally
in itinerant electron systems with randomness
although its effect might be weaker in higher dimensions.
Thus, the anomalies in three-dimensional systems in Fig.~\ref{fig:3D}
may be an analog of the anti-crossing due to the Friedel oscillation.

We discuss our findings in comparison with experimental results.
Our results reproduce the anomalies of spin excitations
reported in low-$T_{\rm C}$ compounds qualitatively:
(i) The spectrum shows a significant broadening near the zone boundary.
\cite{Hwang1998,Vasiliu-Doloc1998,Dai2000,Biotteau2001}
(ii) The softening near the zone boundary
\cite{Hwang1998,Dai2000,Biotteau2001}
can be interpreted
as a consequence of the branching or the anti-crossing in our results.
The lower-energy branch has a substantial weight and
is easily identified in neutron scattering experiments,
while the higher-energy one may be difficult to identify
due to a deficiency of energy resolution in experiments.
(iii) The anti-crossing or the gap opening
\cite{Vasiliu-Doloc1998,Biotteau2001}
have been reproduced.
Thus, the randomness is reasonably considered as a hidden parameter
to give a qualitative description for the anomalous spin dynamics 
in low-$T_{\rm C}$ manganites.

Several other scenarios have also been proposed to explain
the spin excitation spectrum in low-$T_{\rm C}$ manganites.
As an origin of the softening,
some additional elements to the DE mechanism have been examined such as
magnon-orbital
\cite{Furukawa1999b}
and magnon-phonon couplings.
\cite{Khaliullin2000}
Magnetic origin has also been proposed.
\cite{Solovyev1999}
For the origin of the broadening,
magnon-phonon
\cite{Furukawa1999b}
or magnon-electron couplings
\cite{Golosov2000}
have been discussed.
These arguments attribute the origin of the
spectrum change by the A-site substitution to the modification of the 
bandwidth through some additional couplings to 
the spin degrees of freedom.
As discussed previously, however, the actual changes of the bandwidth
in these compounds are small.
Therefore, it seems difficult for these scenarios to explain the
large changes of the spin excitations in a quantitative manner.
Moreover,  neither the anti-crossing nor the gap opening
has been reproduced by them.

Our results give a comprehensive scenario to understand
the systematic changes of spin excitations in the A-site
substituted manganites,
since the change of the randomness are expected to be large,
as previously discussed in reproducing 
the systematic decrease of $T_{\rm C}$.
We suggest that
the randomness gives important effects on various thermodynamic
properties of A-site substituted manganites.

We now propose some experimental tests for our scenario.
One is the sample dependence.
Our scenario predicts a correlation between the sample quality and
the anomalies in the spin wave spectrum.
Therefore, even in compounds with the same chemical formula,
the extent of the anomalies would depend on the purity of samples
which is measured independently, for instance, by the residual resistivity.
The second test is to investigate the location of the anomalies
in the whole Brillouin zone, and
to make a comparison with information of the Fermi surface,
which is obtained by independent experiments,
for instance, by the angle-resolved photoemission.
The anomalies appear at the Fermi wave number in our scenario.
Doping dependence is also crucial in this test.
The third is to observe the high energy excitation
near the band top $\mib{q} \simeq (\pi,\pi,\pi)$.
In our results,
there remains an excitation with substantial weight and
relatively sharp peak-width near the band top
even under strong randomness.
This may be observed by the pulse-neutron scattering.
All these experimental tests are feasible and
may help to discuss the origin of the anomalies in the spin excitation
as well as the role of the A-site substitution in CMR manganites.

Finally, we comment on higher-order corrections in the $1/S$ expansion.
The aim of this Letter is to study the systematic changes of the
spectrum by the A-site substitution of manganites.
In this case, the parameter $1/S$ should be kept fixed, and
the $1/S$ expansion does not contribute to the spectrum changes.
Furthermore, it has been shown that
the contribution of the higher order terms are very small at
$x \simeq 1/3$ where we are interested in,
at least in the pure case.
\cite{Kaplan1997,Wurth1998}
Therefore, we conclude that the lowest order terms in the
$1/S$ expansion is  sufficient  to discuss the randomness
effects as a first step,
although quantitative estimates of the $1/S$ corrections
in the presence of randomness are left for further studies.


To summarize, we have investigated effects of randomness
on the spin excitation spectrum in the double-exchange model.
The spin dynamics in the ground state is calculated by the spin wave approximation
in the lowest order of the $1/S$ expansion.
Randomness induces some anomalies in the spectrum near the Fermi wave number
such as the broadening, the anti-crossing and the gap opening.
We speculate that the origin of the anomalies is
modulation of effective exchange coupling by the Friedel oscillation.
The anomalous spin dynamics in CMR manganites is reproduced qualitatively.
Since the anomalies in experiments become conspicuous
in A-site substituted compounds which have low Curie temperature,
our results suggest that the randomness is an important element
in the A-site substitution
which has previously been understood as the bandwidth control.
We have discussed our results in comparison with other theories
and proposed several crucial experimental tests.

This work is supported by  ``a Grant-in-Aid from
the Ministry of Education, Culture, Sports, Science and Technology''.


\end{document}